# N-Polar AlGaN/GaN Heterojunction Bipolar Transistor

Md Saklain Morshed[1], John Cobbinah[1], Ijaas Mohamed[1], Jonathan Pratt[1], Akilesh Srikanth[1]
Chandan Joishi[1], Siddharth Rajan[1,2]

[1]*Department of Electrical & Computer Engineering, The Ohio State University, Columbus OH 43210, USA*

[2]*Department of Materials Science & Engineering, The Ohio State University, Columbus OH 43210, USA*

*Corresponding author email: morshed.9@osu.edu*

**Abstract—** Heterojunction bipolar transistors based on the III-Nitride systems have significantly lower performance than field effect transistors due to challenges associated with p-type doping of the base. N-polar AlGaN/GaN HBTs can overcome these challenges by polarization-induced holes in the base and base–emitter interface. We experimentally demonstrate the first N-polar AlGaN/GaN heterojunction bipolar transistors (HBTs) grown by plasma-assisted molecular beam epitaxy on GaN substrate without emitter regrowth. The HBTs exhibit a peak current gain of approximately 15, peak differential current gain of approximately 18, and emitter-area-normalized peak current densities exceeding 20 kA/cm² and 66 kA/cm² for emitter areas of 10 × 10 μm² and 3 × 10 μm², respectively. The specific contact resistance for base is estimated 1.32 × $10^{-2}$ Ω·cm². The open-base breakdown voltage is characterized ~63 V for a 300nm collector without dielectric passivation. This work lays the foundation for future development of N-polar III-Nitride HBTs for next-generation RF applications.

## I. INTRODUCTION

GaN-based heterojunction bipolar transistors (HBTs) are promising for next-generation high-power and high-frequency applications[1)–3)]. When compared with AlGaN/GaN HEMTs, they promise several key advantages – they are not susceptible to surface-related dispersion[4)], they have more uniform distribution of current and thermal dissipation[5)], and the breakdown field is supported by a bulk PN junction (in contrast to a metal-semiconductor junction in HEMTs). However, unlike SiGe or III-As,P material systems where bipolar transistors and field effect transistors often have comparable performance, in the III-Nitride system, bipolar transistors significantly underperform relative to field effect transistors.

The root cause for the poor performance of III-Nitride HBTs is well understood. They are limited by the high Mg acceptor ionization energy and low hole mobility which combine to give poor hole concentration, thereby hindering minority-carrier transport[6)–9)], and a resistive base. Polarization engineering in AlGaN/GaN heterostructures offers an alternative by introducing additional mobile holes through polarization-induced charge accumulation[10)–12)]. Earlier studies have theoretically analyzed collector-up HBTs architectures to utilize this effect in Ga polar orientation[10),13)]. However, in collector-up Ga polar HBT, large-area base–emitter overlap can cause excessive junction leakage. Although dielectric interlayers and ion implantation have been proposed to mitigate leakage, they add fabrication complexity. Simulation of N-polar HBTs shows that the use of graded base designs can enable performance that can exceed the state-of-art mm-wave AlGaN/GaN HEMTs in terms of $f_T V_{BR}$ and $f_{MAX}$ [14)].

In this work, we experimentally demonstrate N-polar AlGaN/GaN HBTs that offer these polarization benefits while retaining a conventional emitter-up architecture. Growth, fabrication, and characterization of N-polar AlGaN/GaN heterojunction bipolar transistors (HBTs) by plasma-assisted molecular beam epitaxy (PA-MBE) on GaN substrates have been presented. The HBTs are fabricated without any regrowth, and the specific contact resistance of base contact is estimated to be $1.32 \times 10^{-2}$ Ω·cm². The demonstrated N-polar HBTs exhibit a peak current gain of ~15, peak differential current gain ~18 an emitter-area-normalized current density of 20 kA/cm² and 66 kA/cm² for emitter areas of 10 × 10 μm² and 3 × 10 μm², respectively. The unpassivated device with a 300-nm collector exhibits an open-base breakdown voltage of ~63 V. Finally, through TCAD simulation, due to presence of 2DHG it is predicted that the thinner base in N polar configuration can obtain >100GHz $f_T$ and $f_{MAX}$. These results establish a foundation for the development of N-polar GaN HBTs and highlight polarization engineering as a potential strategy for overcoming key limitations in III-nitride bipolar devices for next generation RF devices.

## II. EPITAXIAL GROWTH AND FABRICATION DETAILS

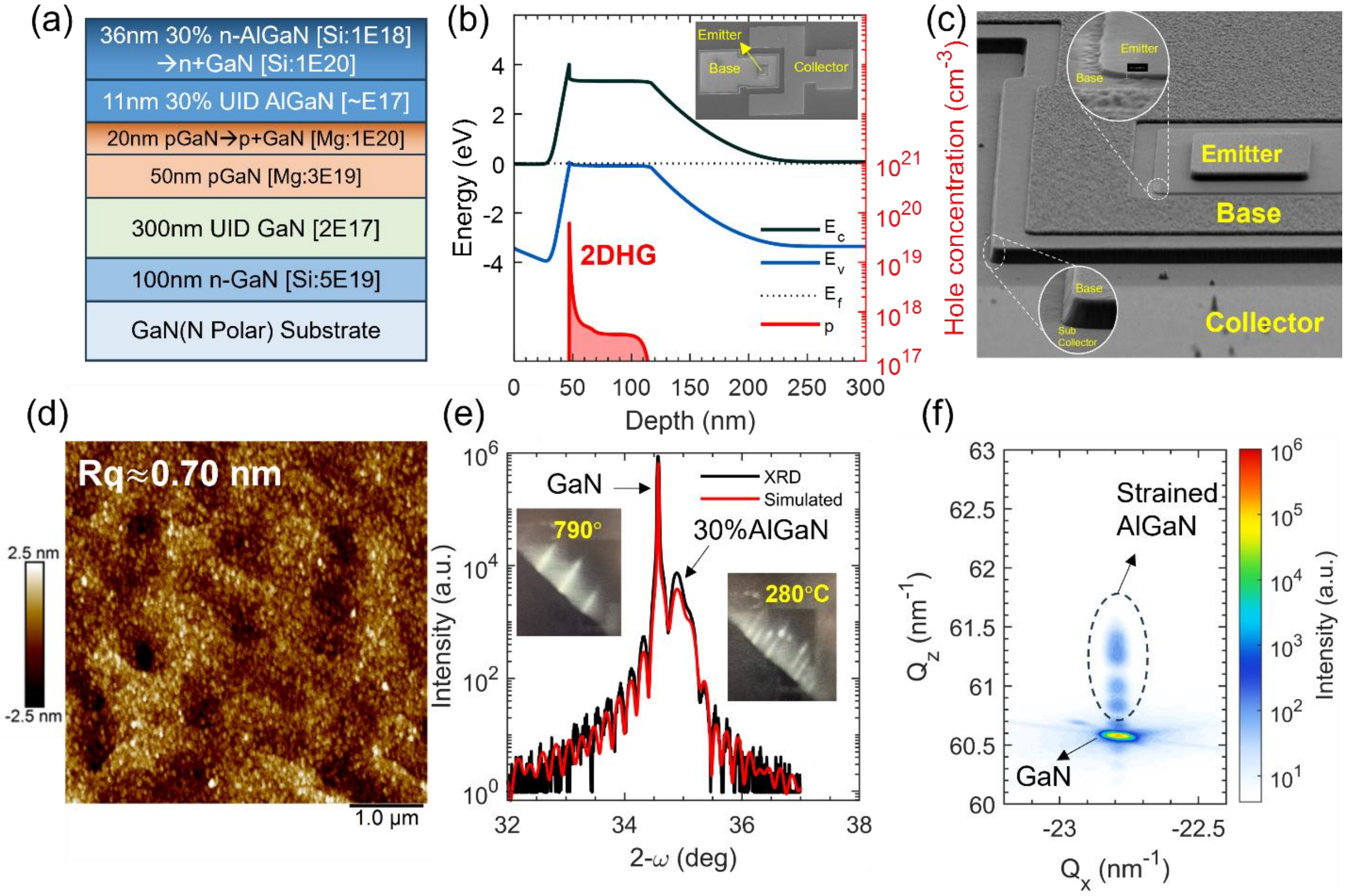

**Fig. 1.** (a) Schematic of the epitaxial structure and (b) corresponding energy-band diagram of the N-polar AlGaN/GaN HBT, illustrating the theoretical formation of a two-dimensional hole gas (2DHG) induced by negative polarization charge at the UID-$Al_{0.3}Ga_{0.7}N$/p-GaN interface. (c) SEM image of the fabricated HBT, showing the emitter, base, and collector metal contacts, along with the emitter and base–collector mesas. (d) 5 × 5 μm² atomic force microscopy (AFM) image of the as-grown epitaxial surface, with an RMS roughness of 0.7 nm. (e) Measured and simulated X-ray diffraction (XRD) patterns, confirming the UID $Al_{0.3}Ga_{0.7}N$ layer (11 nm) and compositionally graded AlGaN layer (36 nm); 3 × 3 RHEED reconstruction at 280°C substrate temperature confirming N polarity after the growth (f) Reciprocal space mapping (RSM) showing that the epitaxial layers are fully strained.

The N-polar AlGaN/GaN HBT epitaxial structure shown in Fig. 1(a) was grown on a single-crystalline N-polar n-GaN bulk substrate (Mitsubishi Chemical America - [15]) with a dislocation density of ~$10^4$ cm$^{-2}$ using a Veeco GEN930 PA-MBE system equipped with an RF $N_2$ plasma source and Ga, Al, Si, and Mg effusion cells. Before growth, the substrate was ultrasonically cleaned in acetone, methanol, and isopropyl alcohol for 5 min each, then In-bonded to a Si wafer. Thermal outgassing was performed at 200 °C for 2 hours in the load-lock and 400 °C for 1 hour in the buffer chamber. Growth was carried out with a background pressure of ~$10^{-5}$ Torr with 1 sccm $N_2$ flow and 300 W RF power.

The growth was performed under metal-rich conditions; Figure 1(d) shows a smooth surface with an RMS roughness of 0.7 nm over a 5 × 5 μm² area. The epitaxial structure was further characterized by high-resolution X-ray diffraction (XRD). Figure 1(e) compares the measured and simulated XRD scans, confirming the presence of an 11-nm UID $Al_{0.3}Ga_{0.7}N$ layer and a 36-nm compositionally graded AlGaN layer. A 3 × 3 reflection high-energy electron diffraction (RHEED) reconstruction was observed at 280 °C after growth, indicating that no polarity inversion occurred during epitaxy. The strained film was evaluated using reciprocal space mapping (RSM), as shown in Fig. 1(f). The RSM results indicate that the epitaxial layers remain fully strained to the GaN substrate.

Device fabrication began with a 47-nm emitter etch using a Plasma-Therm SLR 770 ICP-RIE system. A 420-nm-deep base mesa was then etched using $BCl_3/Cl_2$ plasma. The base contact consisted of Ni (30 nm)/Au (70 nm), annealed at 500 °C for 2 min in $O_2$. Emitter and collector contacts were formed using Ti (30 nm)/Al (120 nm)/Ni (30 nm)/Au (100 nm). The spacing between the base metal to the emitter sidewall and base-mesa to collector metal were maintained at 1 μm.

## III. RESULTS AND DISCUSSION

Figure 2 presents the linear TLM characteristics and junction electrical behavior. Figure 2(a) and (b) show the TLM characteristics of the emitter and base, respectively, using 100 × 30 μm² metal pads. The extracted sheet resistances are 298 Ω/sq for the emitter and 478 kΩ/sq for the base. The corresponding specific contact resistances are $4.75 \times 10^{-7}$ and $1.33 \times 10^{-2}$ Ω·cm², respectively. The collector sheet resistance and specific contact resistance, extracted using cTLM, are 7.5 Ω/sq (thanks to the n-GaN substrate) and $2.69 \times 10^{-6}$ Ω·cm², respectively.

Figure 2(c) and (d) show the electrical characteristics of the base–collector (BC) and base–emitter (BE) junctions, respectively. Both junctions exhibit clear forward conduction and reverse-bias blocking, with on/off current ratios exceeding $10^7$ and $10^4$, respectively. The BC junction yields an on-resistance of $1.32 \times 10^{-2}$ Ω·cm², indicating a major contribution from the base-contact resistance. In contrast, the BE junction has a higher on-resistance of 1.02 Ω·cm², primarily due to lateral current spreading (for the characterized diode, the base-emitter spacing is ~ 5μm) through the high-resistance p-GaN base. The extracted diode ideality factors are 3.24 and 4.16 for the BC and BE junctions, respectively.

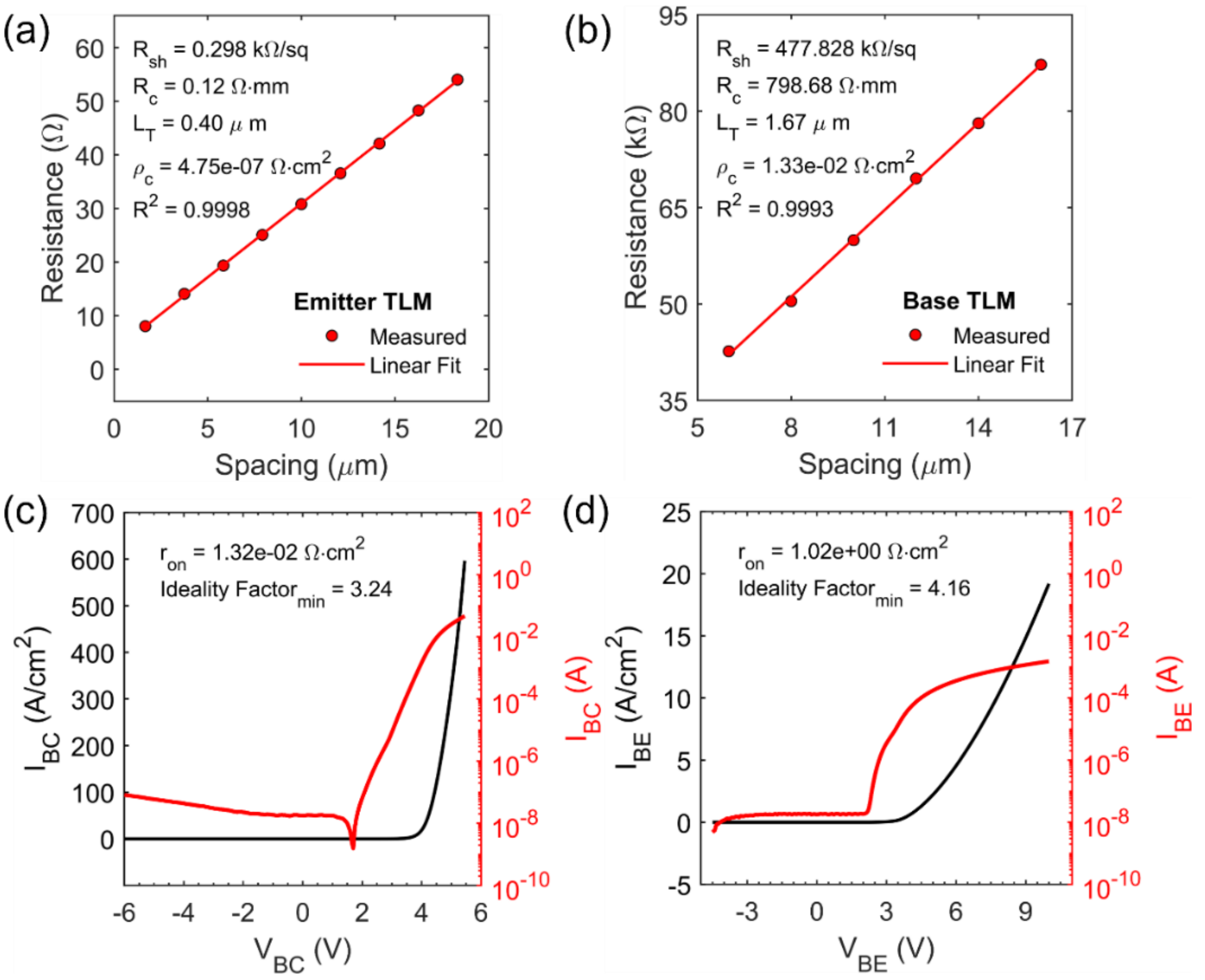

Fig. 2. Linear transfer length method (TLM) characteristics of the (a) emitter and (b) base. Current–voltage characteristics of the (c) base–collector and (d) base–emitter diodes, presented on both linear and logarithmic scales.

The transistor characteristics of the fabricated N-polar AlGaN/GaN HBTs are shown in Figure 3. Figure 3(a) shows the common-emitter Gummel plot of two devices with an emitter area of 10 × 10 µm² and 3 × 10 µm², measured at a fixed collector–emitter voltage of 15 V. For a 10 × 10 µm² HBT, a peak current gain of ~15 is estimated at a collector current of 21.5 mA, corresponding to an emitter-area-normalized current density of 21.5 kA/cm². For a narrower emitter of 3 × 10 µm² the maximum current becomes almost identical; thus, the peak current density exceeds ~66.6 kA/cm². Both current gains and current densities are comparable with MBE-grown Ga polar HBTs[16)–18)]. The peak differential current gain is approximately ~18. The collector and base currents cross over at approximately 3.35 V, which is comparable with other MOCVD works [19)] and lowest in terms of MBE grown HBTs.

Figure 3(b) presents the common-emitter output characteristics of the emitter area of 10 × 10 µm² HBT. The collector–emitter voltage was swept from 0 to 25 V in 0.05 V increments, with the base current varied from 100 to 500 µA in 50 µA increments. The estimated offset voltage is 2.1 V. The relatively high offset voltage may be associated with higher contact resistance resulting from plasma-induced damage during emitter-mesa etching. At a base current of 500 µA, the knee voltage is approximately 10.4 V at a collector current of 5.7 mA. Figure 3(c) depicts the open-base breakdown voltage is around ~63V. Figure 3(d) presents the common-emitter output characteristics measured at higher base currents to investigate self-heating. The base current was swept from 100 to 1500 µA in 200 µA increments. Self-heating becomes pronounced at collector current densities above approximately 10 kA/cm², resulting in deviations from the expected output characteristics at high current levels.

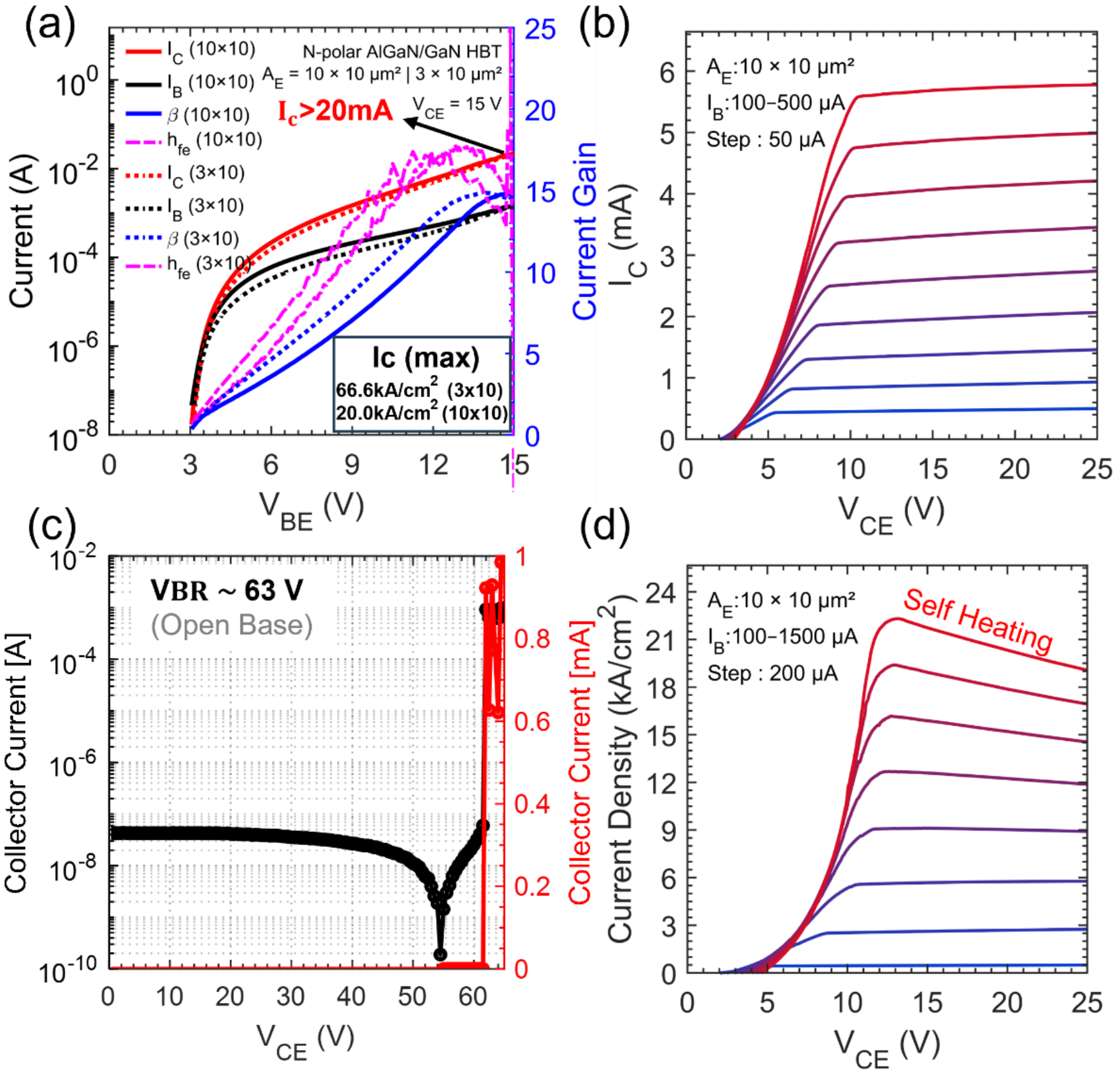

Fig. 3. (a) Common-emitter Gummel plot measured at a fixed collector–emitter voltage ($V_{CE}$) of 15 V for HBT emitter areas of 10 × 10 µm² and 3 × 10 µm². (b) Common-emitter output characteristics measured with $V_{CE}$ swept up to 25 V and base current steps of 50 µA from 100 to 500 µA. (c)Breakdown characteristics of open-base configuration. (d) High-current output characteristics demonstrating device operation (self-heating) at emitter-area-normalized current densities exceeding 20 kA/cm².

## IV. PROSPECTS OF N POLAR AlGaN/GaN HBTs

In this section we present 2D TCAD simulations using Silvaco Atlas to calibrate the simulation framework to our experimentally demonstrated HBT and evaluate its potential for future performance improvements. Initially, the simulated device geometry was kept identical to the fabricated structure and then the ultimate performance limit is evaluated with device scaling(200nm emitter width) and thinner(30nm) base. Figure 4(a) shows the simulated Gummel characteristics, showing good agreement in the overall current gain behavior. The slow increase in current gain is attributed to the degraded base contact resulting from dry etching.

For calibration, we considered electron and hole mobilities of 120 cm²/V·s and 12 cm²/V·s, respectively, and a minority-carrier lifetime of 20 ps[20]. The physics based models include Fermi–Dirac statistics, incomplete ionization, Auger and Shockley–Read–Hall recombination, concentration-dependent mobility, bandgap narrowing, polarization effects, and the relevant material parameter sets. The Mg acceptor activation energy is set to 147 meV, while the donor activation energy is 10 meV. The electron saturation velocity is $2 \times 10^7$ cm/s [21]. The optical recombination coefficient (Copt) is $1.1 \times 10^{-10}$ cm³/s[22], and the Auger recombination coefficient is $10^{-30}$ cm⁶/s [23),24)]for both electrons and holes.

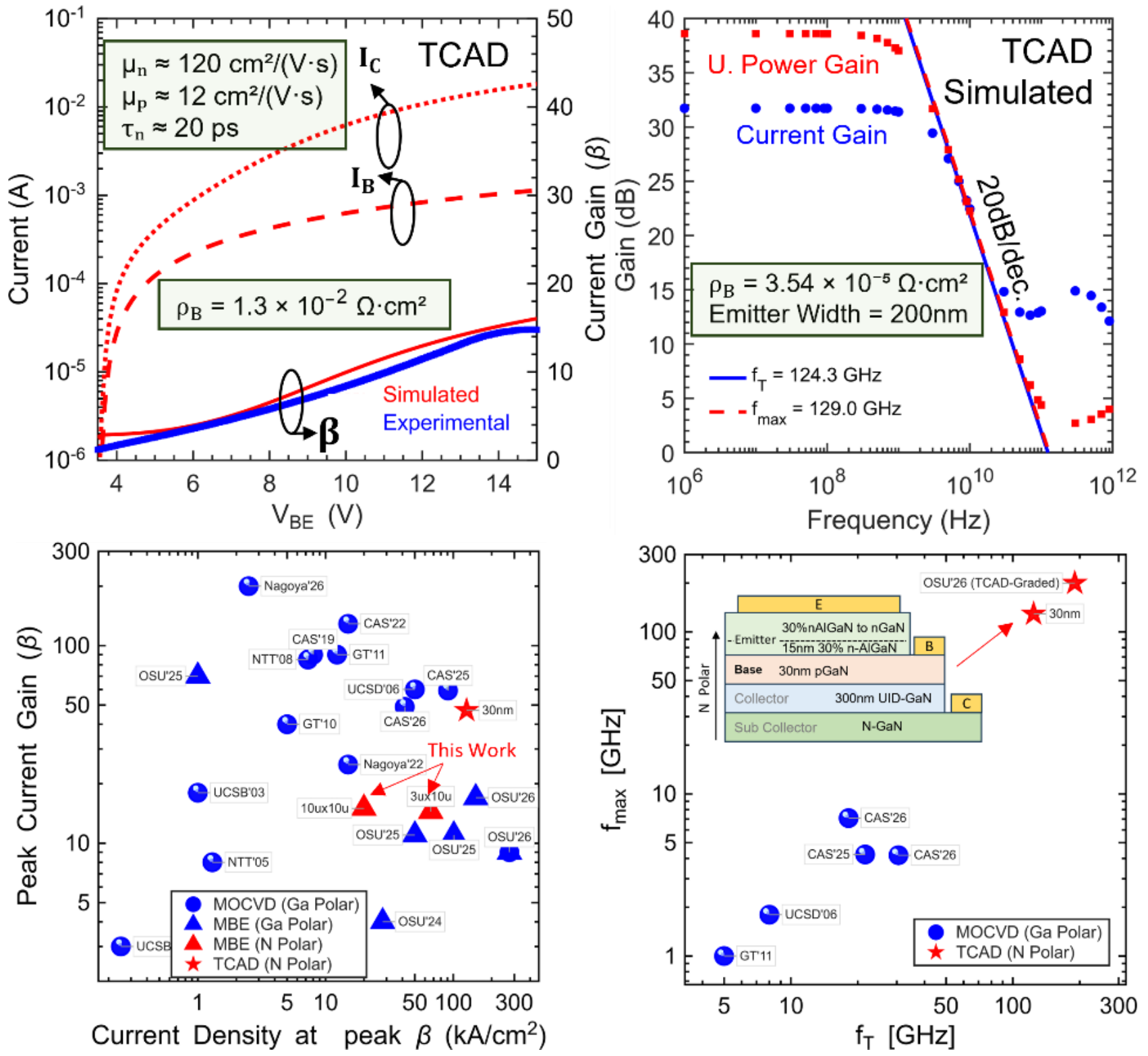

**Fig. 4.** (a) Gummel plot of the TCAD-simulated N-polar AlGaN/GaN HBT compared with experimental data. (b) Small-signal characteristics of the scaled HBT with a 200-nm emitter width and 30-nm base thickness. Benchmark comparisons of (c) peak current gain versus current density at peak gain and (d) maximum oscillation frequency ($f_{MAX}$) versus cutoff frequency ($f_T$) [3),16)–19),25)–33)].

Following DC model calibration, small-signal simulations were performed to evaluate the RF performance of a scaled HBT with a 200-nm emitter width, as shown in Fig. 4(b). A major limitation of conventional Ga-polar AlGaN/GaN HBTs is the low hole concentration achievable in the Mg-doped p-GaN base due to the high Mg acceptor activation energy (150–220 meV). This limits base thinning while maintaining sufficient conductivity. Consequently, conventional devices typically require bases around 70 nm, that increases minority-carrier transit time and recombination, thereby degrading the base transport factor and current gain.

In contrast, the polarization-induced two-dimensional hole gas (2DHG) at the N-polar AlGaN/GaN heterointerface provides additional hole accumulation, enabling a thinner base while maintaining a high sheet hole density. This advantage allows us to investigate an RF HBT with a 30-nm base. The base contact resistance dominates $f_{MAX}$; therefore, the experimentally reported etched p-GaN base contact resistance of $3.54 \times 10^{-5}$ Ω·cm² [34)]was considered into the simulation. The simulated device achieves a cutoff frequency ($f_T$) of 124 GHz and a maximum oscillation frequency ($f_{MAX}$) of 129 GHz. The collector thickness is maintained at 300 nm. Based on the theoretical breakdown field of GaN, a breakdown voltage of approximately 100 V can be considered. The non-graded base lacks a quasi-electric field compared to a graded base, resulting in increased base transit time compared to an pGaN-to-pAlGaN graded base, which is projected to achieve even higher cutoff frequency and maximum oscillation frequency by our another work [14)].

## V. CONCLUSION

In conclusion, N-polar AlGaN/GaN HBTs have been experimentally demonstrated using PA-MBE on GaN substrates without requiring emitter regrowth. The fabricated devices exhibit a peak current gain of ~15 and an emitter-area-normalized current density exceeding 66 kA/cm². TCAD simulations further indicate that polarization-induced hole accumulation can provide an effective approach to improving base conductivity and reducing the base thickness, thereby enhancing DC characteristics. These results also suggest the potential of N-polar HBTs to surpass the performance of existing AlGaN/GaN RF HBTs. Further optimization of epitaxial growth, base contacts, and device scaling will be essential to improve overall transistor performance.

**Acknowledgements:** We acknowledge funding from AFOSR FA9550-26-1B165 (Program Manager Dr. Michael Yakes) and ARO DEVCOM W911NF2220163 (Program Manager Dr. Tom Oder)

**Author Declarations**

**Conflict of Interest:** The authors have no conflicts to disclose.